\documentclass[journal]{IEEEtran}
\usepackage{graphicx}

\ifCLASSINFOpdf
\else
\fi
\usepackage[cmex10]{amsmath}
\usepackage{url}

\IEEEoverridecommandlockouts

\usepackage{tikz}
\usepackage{textcomp}
\usepackage{hyperref}

\newcommand\copyrighttext{
  \footnotesize \textcopyright 2016 IEEE. Personal use of this material is permitted.
  Permission from IEEE must be obtained for all other uses, in any current or future 
  media, including reprinting/republishing this material for advertising or promotional 
  purposes, creating new collective works, for resale or redistribution to servers or 
  lists, or reuse of any copyrighted component of this work in other works. 
  }
\newcommand\copyrightnotice{
\begin{tikzpicture}[remember picture,overlay]
\node[anchor=south,yshift=10pt] at (current page.south) {\fbox{\parbox{\dimexpr\textwidth-\fboxsep-\fboxrule\relax}{\copyrighttext}}};
\end{tikzpicture}
}

\begin{document}
%

\title{POLAR: Final Calibration and In-Flight Performance of a
Dedicated GRB Polarimeter}
%
%
%

\author{M. Kole, T.W. Bao, T. Batsch, T. Bernasconi, F. Cadoux, J.Y. Chai, Y.W. Dong, N. Gauvin, W. Hajdas, J.J. He, M.N. Kong, S.W. Kong, C. Lechanoine-Leluc, L. Li, Z.H. Li, J.T. Liu, X. Liu, R. Marcinkowski, S. Orsi, M. Pohl, N. Produit, D. Rapin, A. Rutczynska, D. Rybka, H.L. Shi, L.M. Song, J.C. Sun, J. Szabelski, R.J. Wang, Y.J. Wang, X. Wen, B.B. Wu, X. Wu, H.L. Xiao, S.L. Xiong, H.H. Xu, M. Xu, J. Zhang, L. Zhang, L.Y. Zhang, S.N. Zhang, X.F. Zhang, Y.J. Zhang, A. Zwolinska

\thanks{Merlin Kole is with the University of Geneva, DPNC, 24, Quai Ernest-Ansermet, CH-1211 Geneve 4, Switzerland, e-mail: merlin.kole@unige.ch}
\thanks{T.W. Bao, J.Y. Chai, Y.W. Dong, J.J. He, M.N. Kong, S.W. Kong, L. Li, Z.H. Li, J.T. Liu, X. Liu, H.L. Shi, L.M. Song, J.C. Sun, R.J. Wang, Y.J. Wang, X. Wen, B.B. Wu, X. Wu, S.L. XIONG, H.H. Xu, M. Xu, J. Zhang, L. Zhang, L.Y. Zhang, S.N. Zhang, X.F. Zhang, Y.J. Zhang are with the Key Laboratory of Particle Astrophysics, Institute of High Energy Physics, Chinese Academy of Sciences, Beijing, China, 100049}
\thanks{T. Batsch, D. Rybka, A. Rutczynska, J. Szabelski and A. Zwolinska are with the National Centre for Nuclear Research. A. Soltana 7, 05-400 Otwock, Swierk, Poland}
\thanks{T. Bernasconi, I. Cernuda, N. Gauvin and N. Produit are with the University of Geneva, ISDC Data center for Astrophysics, 16, Chemin d'Ecogia, 1290 Versoix Switzerland}
\thanks{F. Cadoux, C. Lechanoine-Leluc, S. Orsi, M. Pohl, D. Rapin and X. Wu are with the University of Geneva, DPNC, 24, Quai Ernest-Ansermet, CH-1211 Geneve 4, Switzerland}
\thanks{I. Britvitch, W. Hajdas, R. Marcinkowski and H.L. Xiao are with Paul Scherrer Institut 5232 Villigen PSI, Switzerland}}

\maketitle

\copyrightnotice
\begin{abstract}

Gamma-ray polarimetry is a new powerful tool to study the processes responsible for the emission from astrophysical sources and the environments in which this emission takes place. Few successful polarimetric measurements have however been performed thus far in the gamma-ray energy band due to the difficulties involved. POLAR is a dedicated polarimeter designed to perform high precision measurements of the polarization of the emission from gamma-ray burst in the 50-500 keV energy range. This new polarimeter is expected to detect approximately 50 gamma-ray bursts per year while performing high precision polarization measurements on approximately 10 bursts per year. The instrument was launched into lower earth orbit as part of the second Chinese space lab, the Tiangong-2, on September 15th 2016 and has been taking data successfully since being switched on one week after. The instrument uses a segmented scintillator array consisting of 1600 plastic scintillator bars, read out by 25 flat-panel multi-anode photomultipliers, to measure the Compton scattering angles of incoming photons. The small segmentation and relatively large uniform effective area allow the instrument to measure the polarization of a large number of transient events, such as gamma-ray bursts, with an unprecedented precision during its two year life-time. The final flight model underwent detailed calibration prior to launch as well as intensive space qualification tests, a summary of which will be presented in this paper. The instrument design will be discussed first followed by an overview of the on-ground tests, finally the in-orbit behavior as measured during the first weeks of the mission will be presented.

\end{abstract}

\begin{IEEEkeywords}
Gamma-Ray Burst, Polarimeter, Gamma-ray, X-ray, Instrumentation, Astrophysics.
\end{IEEEkeywords}

%
\IEEEpeerreviewmaketitle

\vspace{2cm}

\section{Introduction}
%
%
%
%
\IEEEPARstart{G}{amma}-Ray Bursts (GRBs) are sudden flashes of gamma-rays which outshine all other gamma-ray sources in the sky for a duration of several seconds. These events have an extra-galactic origin and are considered to be the brightest events in the universe since the Big Bang. In recent years several instruments have performed detailed measurements of the high energy component of GRBs, however much remains unknown about the emission mechanisms and the emission environments involved. Measuring the two polarization parameters of gamma-rays emitted by astrophysical sources such as GRBs is expected to provide new insights in such objects \cite{Toma}, \cite{ChrisL}. In recent years several polarimetric measurements have been performed of GRB emission at high energies, see for example \cite{Yone} and \cite{Yone2}. These measurements drew large attention from the scientific community, however, due to the difficulty in performing such measurements with non-dedicated detectors and the low detection efficiency of polarimetric measurements non of the performed measurements has been able to exclude or put constraints on any of the existing emission models. POLAR is a new joint European-Chinese dedicated GRB polarimetry mission with a relatively high sensitivity and a large effective area. This pioneering instrument was successfully launched on September 15th 2016. One week after launch the instrument commenced data taking. The instrument uses a segmented scintillator array to deduce the the polarization of gamma-rays in the $50-500\,\mathrm{keV}$ energy range by measuring the Compton scattering angle of these photons. In the Compton scattering process a photon is scattered off an electron considered to be at rest. The differential cross section for the process is given by the Klein-Nishina formula \cite{Klein} shown in equation \ref{KN}.

\begin{equation} \label{KN}
\mathrm{\frac{d\sigma}{d\Omega}=\frac{r_0^2}{2}\frac{E'^2}{E^2}\left(\frac{E'}{E}+\frac{E}{E'}-2\sin^2\theta \cos^2\phi\right).}
\end{equation}

Here $\mathrm{E}$ and $\mathrm{E'}$ are the initial and final photon energy respectively, $\mathrm{r_0^2}$ is the classical electron radius, $\theta$ the polar scattering angle and $\phi$ the azimuthal scattering angle with respect to the polarization vector. The $\cos^2\phi$ term results in a polarization dependent cross section which, in turn, results in an increased probability for photons to scatter perpendicular to the polarization vector. The final distribution of the azimuthal scattering angle will be sinusoidal with a period of $180^\circ$, referred to as the modulation curve. The relative amplitude, the amplitude divided by the mean, of this is linearly proportional to the polarization degree through $\mu=\mu_{100}\,\mathrm{P}$. Here $\mu$ is the relative amplitude extracted from the modulation curve, $\mathrm{P}$ is the polarization degree of the measured photon flux and $\mu_{100}$ is the relative amplitude for a $100\%$ polarized beam. The $\mu_{100}$ parameter is an energy dependent instrument specific value which needs to be acquired through calibration with a polarized beam or through detailed Monte Carlo simulations.

The POLAR detector \cite{Nicolas} is designed to measure the Compton scattering angle of incoming photons using a low-Z target of 1600 plastic scintillator bars, divided into 25 modular units. Each module is read out by a Hamamatsu H8500, a flat-panel 64 channel multi-anode photomultiplier (MAPMT). Due the fine segmentation of the scintillator target the Compton scattering angle can be measured with a relatively high precision, resulting in a $\mu_{100}$ of around $40\%$ at $140\,\mathrm{keV}$ which translates into a high sensitivity to polarization. The use of a full plastic array results in a high Compton scattering cross section for photons in the 10's of keV energy range, thereby making the instrument sensitive to the lower energy component of the GRB gamma-ray emission. Lastly the uniformity of the detection surface results in a relatively large effective area and a large field of view which covers approximately 1/3 of the sky, making the instrument optimal for performing polarimetric measurements on transient events such as GRBs \cite{Estel}. 

\begin{figure}
  \begin{center}
    \includegraphics[width=0.55\textwidth]{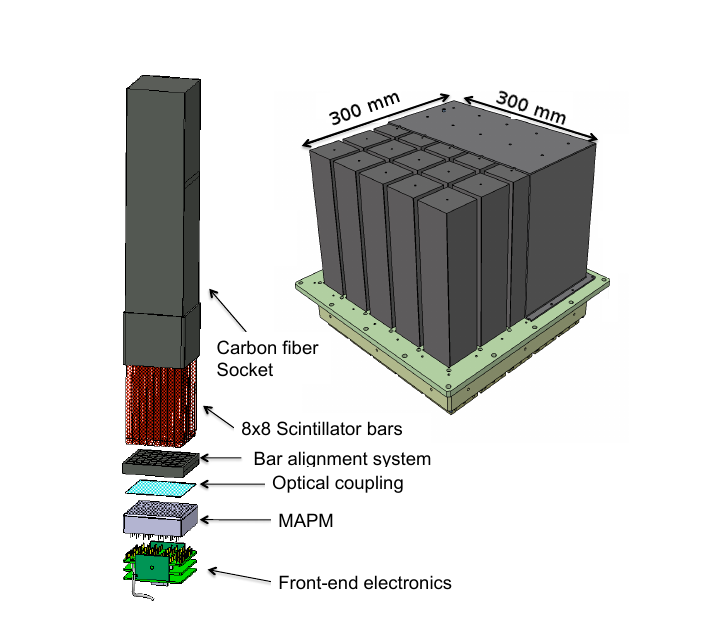}
  \end{center}
  \caption{Schematic overview of one single POLAR module together with the full 5x5 module instrument \cite{Estel}.}
  \label{module}
\end{figure}

\begin{figure}
  \begin{center}
    \includegraphics[width=0.40\textwidth]{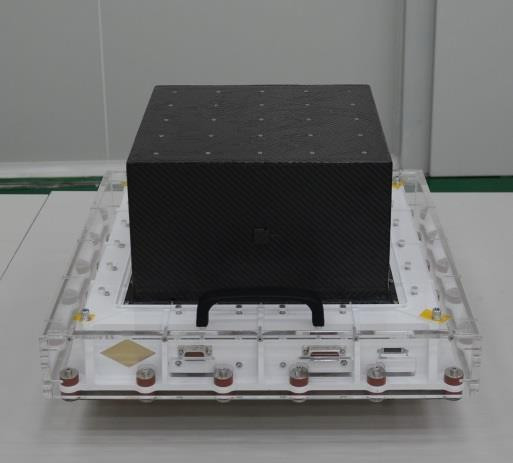}
  \end{center}
  \caption{The full flight ready POLAR instrument.}
  \label{FM}
\end{figure}

The scintillator bars used in POLAR have a surface area of 6 by 6  mm and a length of 176 mm. Optical cross talk between the scintillators is reduced both by placing a reflective vikuiti foil \cite{vikuiti} between the bars and by shaping the top and bottom of the scintillator into a conical shape in which the surface is reduced to 5 by 5 mm. The scintillators are coupled to the MAPMT using an optically transparent pad consisting of Mapsil QS1123 \cite{Mapsil}. This material provides a good optical coupling to the MAPMT and absorbs vibrations and shocks which could cause damage to the PMT window. For further shock protection a 3 mm thick layer of rubber is placed at the top of the scintillators. The 64 bars which make up a module are surrounded by a 1 mm thick layer of carbon fiber for mechanical support. Each MAPMT is read out using its own front-end electronics (FEEs), the 25 FEEs are read out by 3 central FPGAs.  The full effective area of the instrument is covered by a 1 mm thick layer of carbon fiber for mechanical stability. The outside layer is placed for mechanical support while also serving as a passive shield for low energy charged particles. Additionally the full instrument is covered by a multi-layer insulator (MLI) for thermal stability. A schematic overview of a single module together with the full instrument can be seen in figure \ref{module}. The full flight ready detector, without the MLI cover, can be seen in figure \ref{FM}.

The measurement of the Compton scattering angle requires the detection of two interactions of each photon inside the scintillator array. The instrument therefore triggers on events consisting of a minimum of 2 channels being above an adjustable threshold level. Each FEE counts the number of triggering channels within the module and communicates this with the central FPGAs. All 64 channels in the module are read out in case this module has more than 2 triggering channels, or if the module has 1 triggering channel and at least one other module has a coincident triggering channel. Modules without triggering channels are not read out to reduce the data size of each event. The timing precision of trigger event is of the $\mathrm{\mu s}$ level.

In order to reduce the background from charged cosmic rays, which typically leave a track consisting of many triggered channels in the instrument, a veto is issued in case the number of triggering channels inside a module exceeds a set limit. As photon-induced high energy depositions in one channel can induce triggers in neighboring channels through cross-talk the maximum multiplicity is conservatively set to 10 channels. A second veto against cosmic rays is based on the total energy deposition within the module as charged cosmic rays typically deposit several MeV within several channels. Both the level of the total energy deposition and the maximum multiplicity are parameters in the firmware which can be changed during the mission. In order to be able to study the validity of the cosmic ray veto offline 1 in 16 of the vetoed events are stored in the data. Similarly 1 in 256 of all the single multiplicity events are stored for offline studies.

\section{Instrument Calibration}

\subsection{Space Qualification Tests}

The flight model of POLAR underwent extensive space qualification tests in the summer of 2015 at the SERMS test facility in Terni, Italy. The instrument was subjected to random vibrations in the 10-2000 Hz frequency with a root mean square acceleration of 6.68 $\mathrm{G}_{rms}$ for a duration of 60 seconds on all 3 axes and sinusoidal vibration tests in the 4-100 Hz frequency range with amplitudes in the order of 5 g for the same duration. The instrument furthermore underwent shock tests up to 180 g on all 3 axes and a week long thermal vacuum test in the temperature range of -40 to +40 $^\circ$C. The POLAR flight model underwent all tests without any issues.

\subsection{Polarized Beam Test}

\begin{figure}
  \begin{center}
    \includegraphics[width=0.50\textwidth]{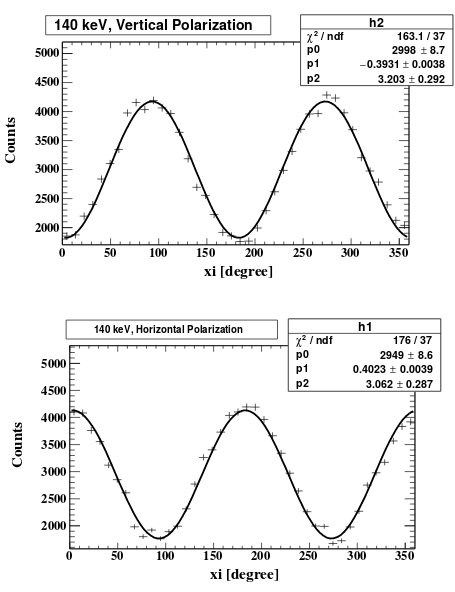}
  \end{center}
  \caption{Preliminary modulation curves as measured by POLAR for a 140 keV on-axis beam for two perpendicular polarization directions, taken from \cite{Hualin}.}
  \label{Mod}
\end{figure}

The flight model of the POLAR detector was calibrated and tested extensively in the two years prior to its launch using both radio active sources and dedicated beam tests. The response of the instrument to polarized photons was tested using a polarized beam at the European Synchrotron Research Facility (ESRF) in Grenoble, France in May 2015. The flight instrument was irradiated using a $100\%$ polarized pencil beams with energies of $60,80,100$ and $140\,\mathrm{keV}$. The beam was moved through the center of each bar consecutively thereby effectively irradiating the full instrument. As POLAR is designed to measure transient events appearing at random position in the sky the majority of the photons will enter the instrument off-axis. In order to study the instrument to off-axis polarized photons all measurements at ESRF were repeated for different incoming angles of the beam. The final results of this test will be used to verify Monte Carlo simulations which in turn are required to study the instrument response to background and to different GRBs. The modulation curves acquired with a 140 keV on-axis beam, taken from \cite{Hualin}, are shown in figure \ref{Mod}. These first results of these beam tests, presented in more detail in \cite{Hualin}, show a $\mu_{100}$ value of approximately $40\%$ for a beam energy of $140\,\mathrm{keV}$ for both polarization directions. This is in good agreement MC simulation results. More detailed studies are still ongoing and will be published in the near future. 

\subsection{Flight Optimization}

As polarization measurements are based on measuring asymmetries in the distribution of events within the instrument a highly uniform instrument response is important. POLAR has a total of 1600 channels read out by 25 MAPMTs, each of which is in turn read out by its own FEE. Non-uniformity is induced by differences in the gain of each channel which consists of a combination of differences in light-yield per scintillator, non-uniform PMT response and non-uniform FEE response. Differences in the position of the low energy thresholds for different channels induce an additional source of non-uniform response. In a MAPMT the same HV is applied to each channel, however typical relative gain differences of $50\%$ are found within a single MAPMT. In order to minimize the gain difference induced by the MAPMTs, only the MAPMTs with the lowest spread in gain were selected for the flight model. Furthermore scintillator bars with a lower light yield were coupled to PMT channels with higher gain and vice versa. It was found that the differences in gain per channel induced by the FEE, which can be measured throughout the mission by injecting known charges into each channel, are negligible compared to light yield and MAPMT induced gain differences.  

To further improve the instrument uniformity the low energy threshold of each FE can be set allowing for equalization of the mean threshold positions of the different modules. Furthermore within the module offsets can be applied to the thresholds of each channel allowing for a reduction of the spread of the threshold positions within the module. 

\begin{figure}
  \begin{center}
    \includegraphics[width=0.40\textwidth]{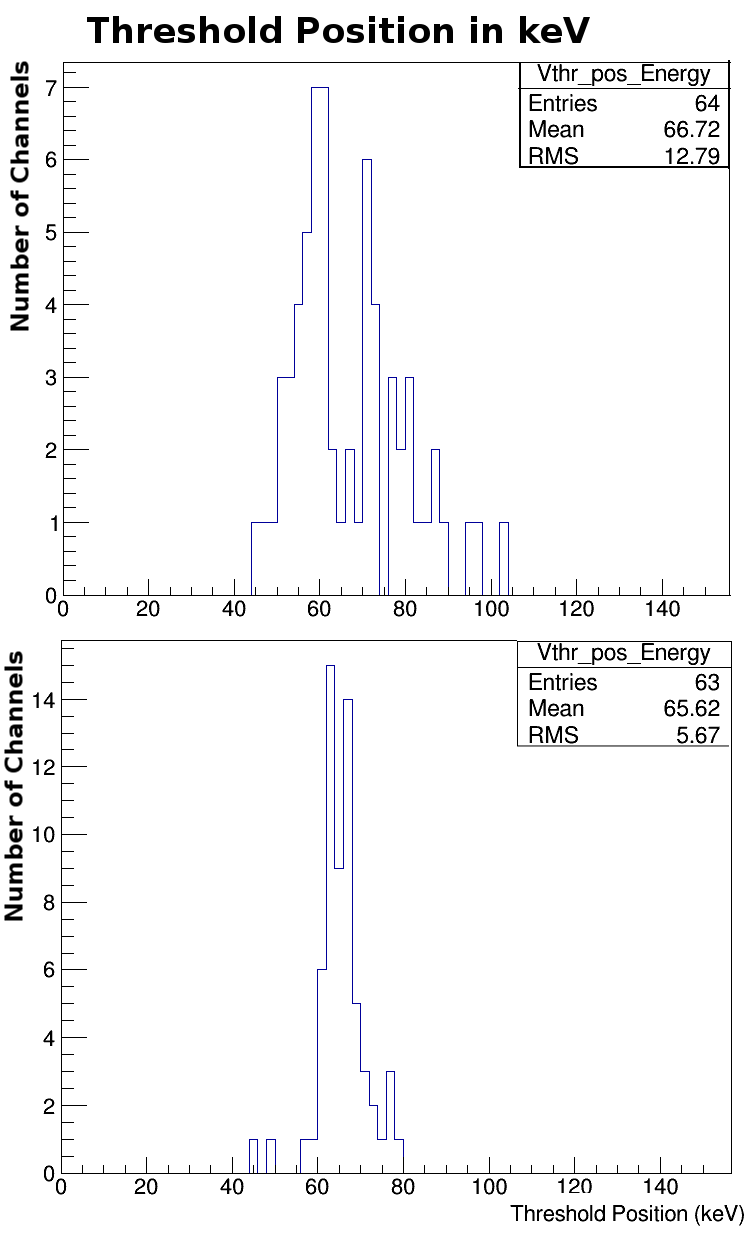}
  \end{center}
  \caption{The position of the lower energy thresholds of a single module in POLAR as measured in keV before optimization (top figure) and after optimization (bottom figure). It should be noted that the threshold positions shown here are not realistic for the flight settings for which the mean is around $10\,\mathrm{keV}$.}
  \label{offsets}
\end{figure}

The dependence of the gain of each channel in POLAR on the applied High Voltage (HV) was calibrated using photons from a Cs-137 source. The results of this calibration in combination with measurements of the dependence of the threshold position of each channel on the applied threshold setting of the FEE makes it possible to reduce the spread in threshold location, as measured in keV, using offsets to a minimum. An example of the effect of the application of threshold offsets can be seen in figure \ref{offsets}. This figure shows that non-uniformity is reduced by more than a factor two resulting in a relative spread below $10\%$. 

\section{In-Flight Performance}

POLAR was launched from the Jiuquan Satellite Launching Center in Inner-Mongolia, China, on the 15th of September 2016 at 22:04 local time as part of the second Chinese Space Lab, the Tiangong-2. Tiangong-2 was placed in a stable lower earth orbit with an inclination of $42^\circ$ at an altitude of approximately $390\,\mathrm{km}$. A heater placed inside the POLAR detector was switched on several hours after launch, keeping the temperature inside the instrument stable around $0^\circ$C. The detector itself was fully switched on on September 23rd at 19:00 Beijing Time after which normal data taking commenced. After switching on all electronics the temperature inside the instrument increased to a mean of $25^\circ$C. Typical temperature fluctuations during an orbit are of the order of $5^\circ$C, the exact values depend on the position of the Sun with respect to the instrument. Since the 23rd of September the instrument has been taking data continuously with exception of a 1 month period, lasting from the the end of October to the end of November, during which Tiangong-2 was manned. During passages through the South Atlantic Anomaly the high voltage of the MAPMTs is generally switched off, all other electronics remains powered on during these passaged. At the time of writing the calibration phase of the instrument is nearing completion and the scientific data taking phase is expected to commence soon.

\subsection{Noise}

\begin{figure}
  \begin{center}
    \includegraphics[width=0.50\textwidth]{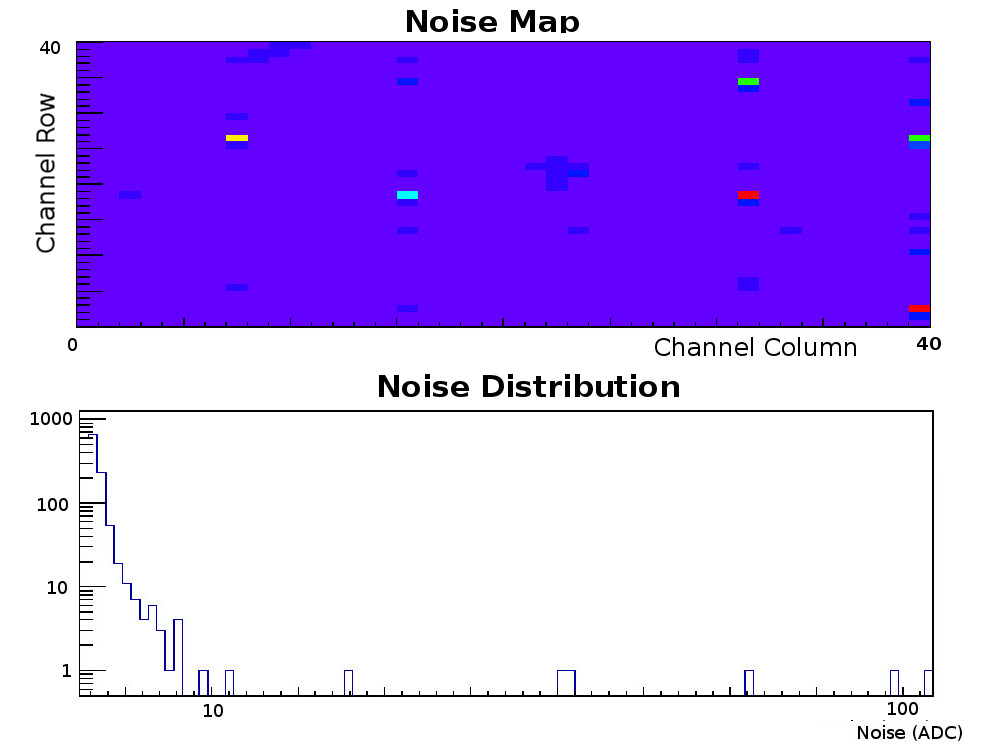}
  \end{center}
  \caption{A overview map of the intrinsic noise levels in each bar of POLAR as a function of their position in the instrument (top) and a histogram of the RMS values of the intrinsic noise levels of the 1600 channels (bottom).}
  \label{noise}
\end{figure}

After launch all of the 1600 channels of POLAR showed normal behavior indicating that all MAPMTs survived the launch and all electronics remained fully functional. The noise as measured in each channel of POLAR consists of two separate components. First each module shows correlated fluctuations of all 64 channels, this noise component is referred to as the common noise. On top of the common noise all channels have individual noise referred to as intrinsic noise. After launch it was found that the common noise levels of each module were fully consistent with the levels measured on ground, indicating no new sources of noise. The intrinsic noise levels changed from channel to channel however the overall distribution of the intrinsic noise levels remained equal to that as measured on ground. The distribution and a map of the noise levels as measured in-orbit can be seen in figure \ref{noise}. With the exception of 7 channels the RMS levels of the intrinsic noise are below 10 ADC, which during normal operation corresponding to $\approx 0.5\,\mathrm{keV}$ when converted to energy. The highest RMS is at 105 ADC. Currently instrument optimization is ongoing. As part of this effort the earlier discussed threshold offsets are being applied to lower the event rate from channels with high noise levels. It is furthermore currently under investigation, using Monte Carlo software, at which noise level it becomes beneficial for polarization measurements to switch a channel off.

\subsection{In-orbit Calibration}

\begin{figure}
  \begin{center}
    \includegraphics[width=0.50\textwidth]{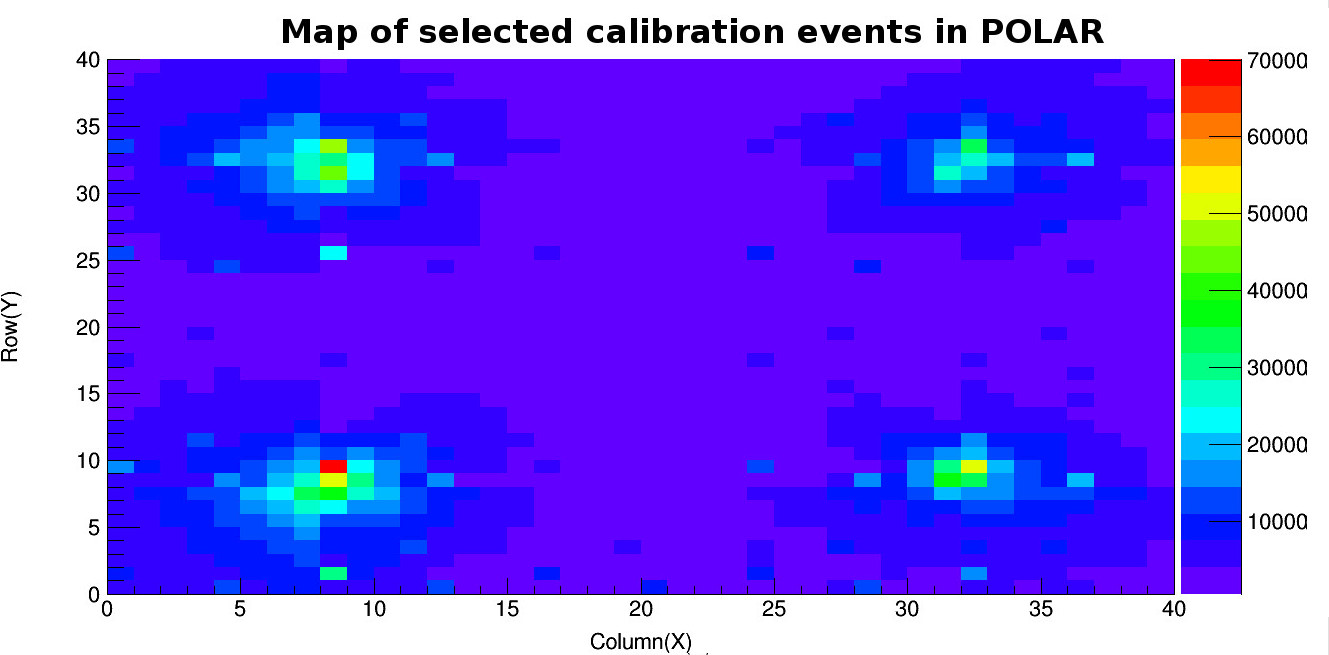}
  \end{center}
  \caption{The number of events per channel after applying the calibration event selection as a function of the channel position in the instrument. The position of the four internal sources is clearly visible.}
  \label{hitmap}
\end{figure}

\begin{figure}
  \begin{center}
    \includegraphics[width=0.50\textwidth]{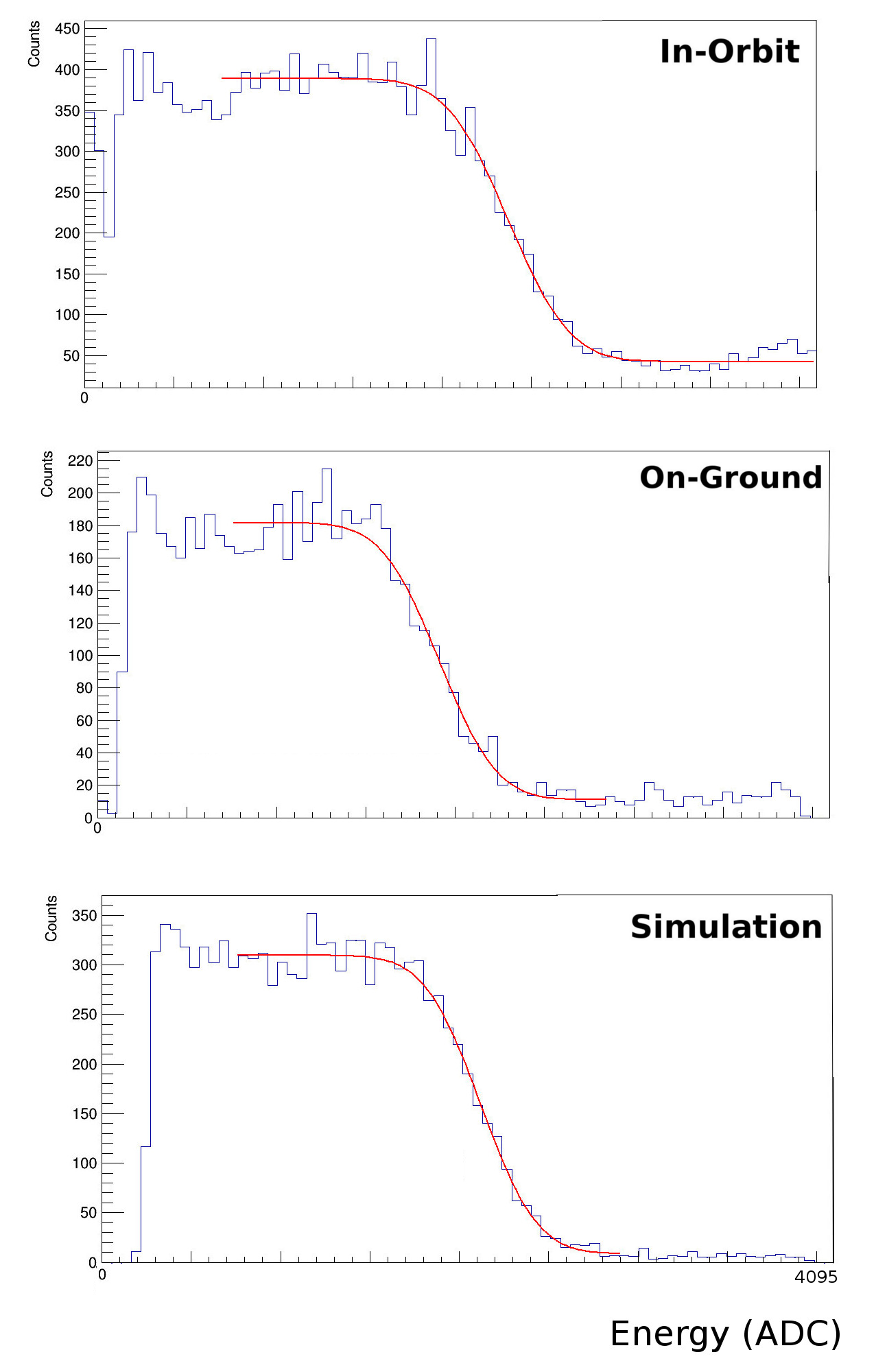}
  \end{center}
  \caption{The energy spectrum after applying the calibration event selection for a single channel as measured in-orbit (top), on-ground (middle) and acquired through simulations (bottom). The Compton edge of 511 keV photons, corresponding to 340 keV, is clearly visible. The difference in the Compton Edge position seen between in-orbit and on-ground measurements is fully compatible with the temperature difference of the instrument for the two measurements.}
  \label{spectra}
\end{figure}

In order to calibrate the energy response of the instrument while in-orbit a total of 4 $\mathrm{Na_{22}}$ sources, each with an activity of approximately 750 Bq, are placed inside of the detector. $\mathrm{Na_{22}}$ emits two back-to-back photons each with an energy of $511\,\mathrm{keV}$. In the event selection the characteristic of back-to-back emission is used together with the knowledge of the positions of the internal sources to apply geometrical cuts. These cuts allow one to select only photons coming from these calibration sources. By selecting only such photons a spectrum can be produced for each channel which shows the characteristic Compton Edge (CE) of the 511 keV photons which is found around 340 keV. Despite the relatively high background rate encountered in-orbit the energy response of the full detector can be calibrated using this method using approximately 12 hours of data. The number of events per channel selected through this method is shown for each channel in figure \ref{hitmap}, the position of the four sources is clearly visible in this figure. Additionally typical calibration spectra of a single channel as measured on-ground, in-orbit and acquired through simulations are shown in figure \ref{spectra}. It can be seen that the position of the CE in the spectrum as measured in-orbit is higher than that measured on-ground. The difference in the position is fully compatible with the difference in the temperature for these two measurements, the in-orbit temperature was approximately $10^\circ$C lower than that during the on-ground measurement. The dependence of the gain on temperature of each channel was measured prior to launch for each channel individually. When comparing the gain of each channel as measured in-orbit to that measured on-ground no significant changes were found. This indicates that the MAPMT behavior did not change as a result of the vibration and shock endured during the launch procedure. This is compatible with earlier measurements which showed no significant change in the gain before and after realistic shock and vibration tests performed at SERMS \cite{Xiaofeng}. A more detailed report on the in-orbit calibration procedure will be published separately in the future.

\subsection{Background Trigger Rates}

\begin{figure}
  \begin{center}
    \includegraphics[width=0.50\textwidth]{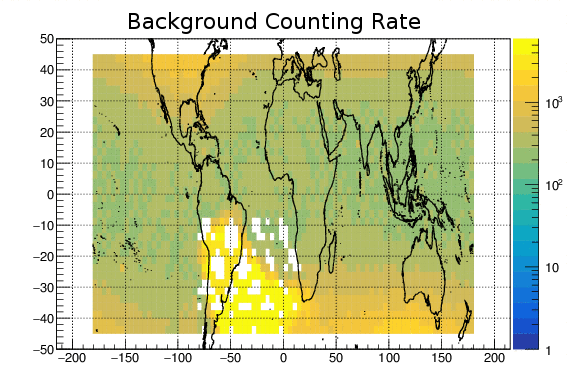}
  \end{center}
  \caption{The trigger rate of the instrument as a function of the position with respect to Earth.}
  \label{rates}
\end{figure}

Unlike other dedicated GRB detectors such as GAP \cite{Yone} POLAR does not require a trigger to start taking data but rather takes data continuously. This is possible due to the relatively large data size reserved for POLAR on Tiangong-2 of 40 GB/day. The background rate as a function of position and time can therefore be studied continuously. The trigger rate as a function of position with respect to the Earth can be seen in figure \ref{rates}. The highest trigger rates were measured during several orbits during which the high voltages were not switched off when passing through the SAA. Outside of the SAA a clear correlation between the trigger rate and the magnetic latitude is observed resulting in a higher trigger rate near the magnetic poles. The position of the trigger distribution within the instrument is furthermore uniform indicating a background which is dominated by photon-like events. A charged particle induced background would result in a higher trigger rate in the peripheral bars. Such a charged cosmic ray induced trigger distribution is only observed inside of the the SAA. These results are therefore in agreement with previous Monte Carlo based background studies predicting a background dominated by the Cosmic X-ray Background (CXB) and atmospheric positrons annihilating in surrounding dead material \cite{Estel}. Both of these background sources are photon-like and the positron induced component is expected to vary with the magnetic latitude \cite{Estel}.

\subsection{GRB Detection}

\begin{figure}
  \begin{center}
    \includegraphics[width=0.50\textwidth]{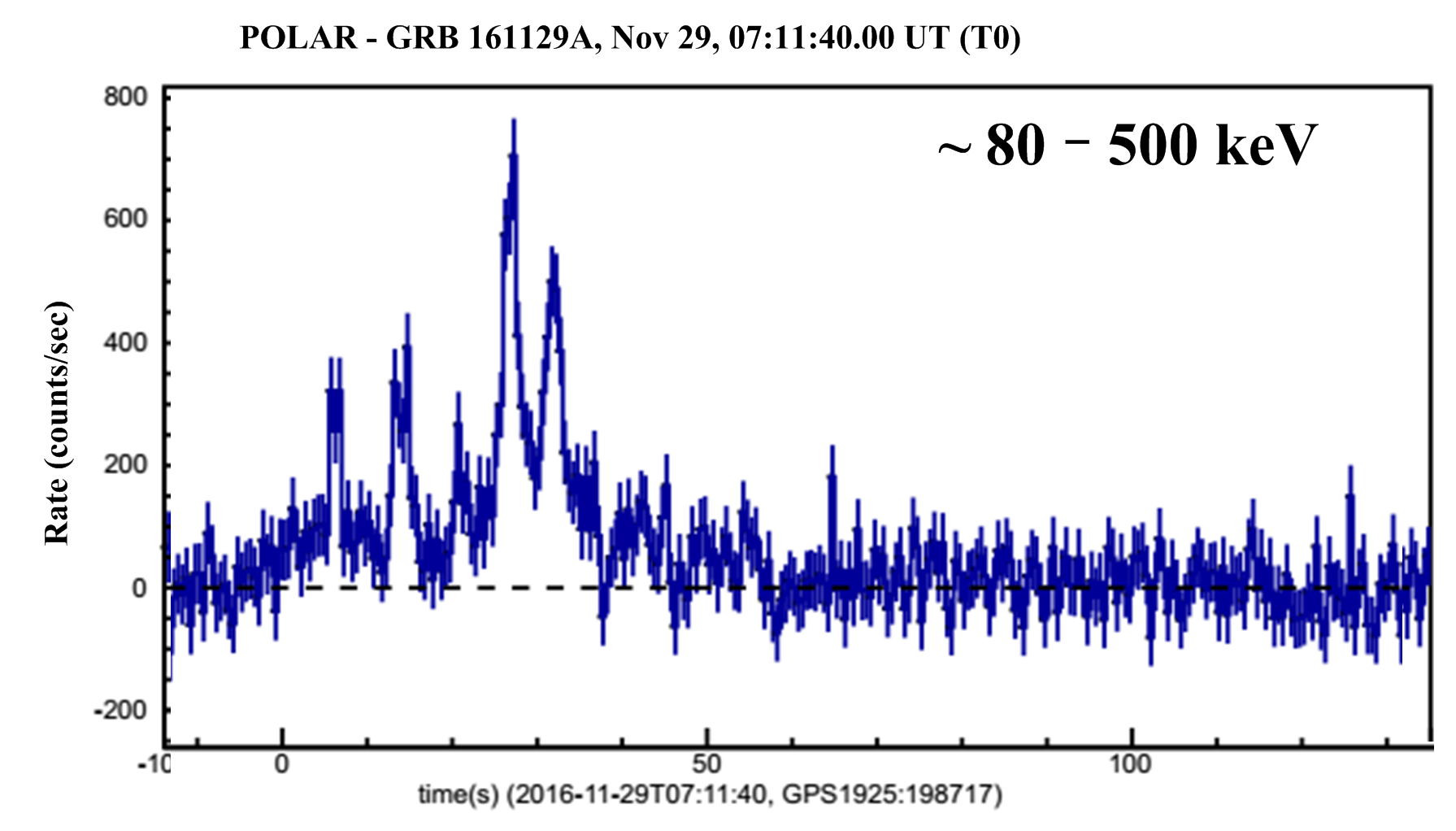}
  \end{center}
  \caption{The light curve as measured by POLAR of GRB 161129A \cite{ISDC}}
  \label{161129A}
\end{figure}

\begin{figure}
  \begin{center}
    \includegraphics[width=0.50\textwidth]{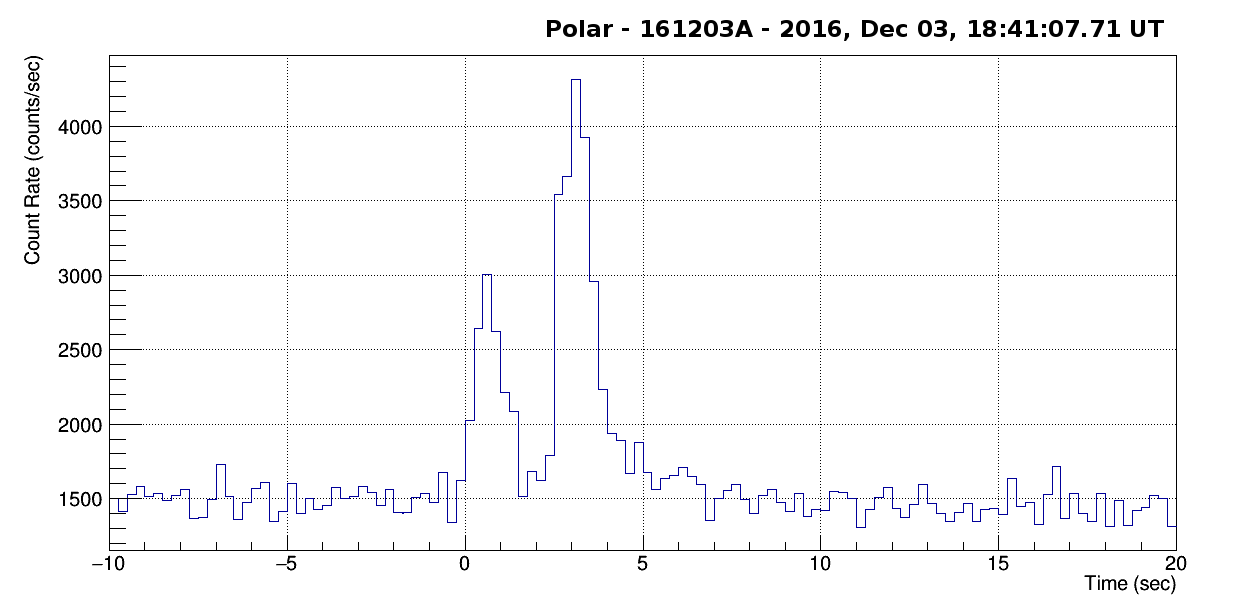}
  \end{center}
  \caption{The light curve as measured by POLAR of GRB 161203A \cite{ISDC}}
  \label{161203A}
\end{figure}

Since data taking commenced POLAR has detected several GRBs, all of which have been measured by other instruments such as Fermi-GBM \cite{Fermiex} and CALET \cite{CALET}. The two most prominent GRBs detected by POLAR as off December 9th 2016, are 161129A and 161203A the light curves of which are shown in figures \ref{161129A} and \ref{161203A} respectively. A GCN circular has been submitted for both these bursts as well as for 160928A. The signal and background rates of these bursts indicate that polarization measurements are possible, however a continuation of studies of the instrument response are required before commencing detailed polarization studies. The reported GRB detections occurred while the mission is still in the calibration phase during which several parameters are being fine-tuned. This phase is currently nearing completion and will be followed by the science data taking phase during which a larger number of GRBs is expected to be measured. Current data analysis focuses only on GRBs confirmed and measured by other instruments allowing for a further calibration of the instrument, after this phase analysis will also focus on detecting bursts which are not measured by other instruments.

\section{Conclusion}

POLAR is a dedicated GRB polarimeter launched as part of Tiangong-2 on September 15th 2016. The instrument is designed to measure the polarization of the GRB emission in the 50-500 keV energy range for approximately 10 GRBs per year. The instrument was switched on on the 23rd of September 2016 after which it was confirmed that POLAR is behaving as expected. It was found that no damage was sustained during launch and no significant changes in the gain of the MAPMTs was detected with respect to on-ground data. The noise levels of the instrument are compatible with those measured on ground, the background as measured in orbit is furthermore compatible with previous Monte Carlo based simulations. Despite still being in the calibration phase the instrument has observed several GRBs, as of the 9th of December 2019 3 GCN circulars were submitted concerning GRB measurements.



%

\ifCLASSOPTIONcaptionsoff
  \newpage
\fi

\end{document}